\documentclass[letterpaper, 12pt]{revtex4}
\usepackage{amsmath}
\usepackage{graphicx}
\begin{document}
\title{Analytic response theory for the density matrix renormalisation group}
\author{Jonathan J. Dorando, Johannes Hachmann, and Garnet Kin-Lic Chan}
\affiliation{Department of Chemistry and Chemical Biology, Cornell University, Ithaca, New York 14853-1301}
\begin{abstract}
We propose an analytic response theory for the density matrix renormalisation group
whereby response properties correspond to analytic derivatives of  density matrix renormalisation group
observables with respect to the applied perturbations. Both static and frequency-dependent response theories
are formulated and implemented. We evaluate our pilot implementation by calculating static and frequency-dependent polarisabilities
of short oligo-di-acetylenes. The analytic response theory
is competitive with  dynamical density matrix renormalisation group methods and
 yields significantly improved accuracies when using a small number of 
density matrix renormalisation group states.
Strengths and weaknesses of the analytic approach are discussed.
\end{abstract}
\date{\today}
\maketitle

\section{Introduction}

The density matrix renormalisation group method \cite{dmrg-original} is now established as a powerful tool 
for ``difficult'' electronic structure
problems in physics and chemistry \cite{SCHOLLWOCK:2005:_dmrg, White1999, Chan2002, chan-oo}. 
In molecular systems, it has been used to describe
multireference correlation in medium-sized active spaces (20-30 active orbitals) for small molecules with complex bonding 
\cite{dmrg-reiher-tm,chan2004b,yanai-cr2, zgid-oo}, as well as a local multireference correlation method in extended long-chain molecules, e.g. to describe excited states in conjugated molecules, using large active spaces  of up to 100 active orbitals \cite{Hachmann2007}.

Response properties, which represent the change in an observable 
as a function of an applied perturbation,  are  of interest in many physical and chemical applications. For example,
geometry optimisation and  vibrational frequencies both require the response of the energy with respect to
changes in the nuclear coordinates, quantities usually known as nuclear derivatives. Nuclear derivatives are examples of 
static response properties because
the perturbation does not depend on time. It is also common to consider frequency-dependent (i.e dynamical) 
response properties
where the applied perturbation is a function of time. 
The most common time-dependent perturbations are fluctuating electric
and/or magnetic fields. In extended systems, the frequency dependence of the response gives insight
into the elementary excitations of the system and this can be used characterise the  nature of the 
electronic ground-state  \cite{kotliar2006esc}.

In many electronic structure methods, response properties are obtained by so-called ``analytic''  techniques.
Analytic response theories of this kind at linear and higher orders have been developed and 
implemented for most  electronic structure methods, including
 Hartree-Fock \cite{mcweeny}, density functional \cite{casida-tddft}, coupled cluster \cite{koch1990eec}, 
multi-configurational self-consistent \cite{olsen1985lan}, and
Moller-Plesset perturbation theories \cite{kobayashi1999cfd}. A review of the formal theory and some of these 
developments may be found in Ref. \cite{olsen-review}.
The name ``analytic'' is used because
the response properties evaluated (e.g. the perturbed energies) correspond strictly 
to derivatives of the ground-state energies or quasi-energies \cite{rice1991cfd, sasagane1993hor, olsen-review}
evaluated in the presence of the perturbation, using  the same level of approximation for the
(quasi-)energy with and without the perturbation.

In contrast, response properties in the density matrix renormalisation group have typically been obtained using a 
quite different approach that appears natural within the DMRG. 
In the DMRG, the wavefunction is  expanded in a set of many-electron
states that are adapted to the state of interest. To obtain a response property, one can choose to solve response equations
using basis states that are  adapted not only to the zeroth order state but also to the calculation of the state's response. These response 
methods, which have
proven very useful in the calculation of dynamical response in DMRG model Hamiltonian calculations, go by the name of Lanczos-vector DMRG \cite{hallberg1995dma}, correction-vector DMRG \cite{ramasesha1996sdm, kuhner1999dcf}, and dynamical
DMRG \cite{jeckelmann2002ddm}. More recently, explicit real-time propagation of the DMRG equations has also been used to obtain 
high-frequency response properties \cite{rt-dmrg}.
A  recent review of all these DMRG response methods can be found in Ref. \cite{jeckelmann-review}. 

In the current work we return to an analytic formulation of response theory  within the density matrix renormalisation group,
in a way that parallels the description of response properties in other electronic structure methods.  We use
as our starting point the wavefunction based (matrix-product state) formulation of the DMRG \cite{Rommer1997, SCHOLLWOCK:2005:_dmrg,chan-oo, zgid-twobody}.
As we shall see, the analytic response approach has a number
of strengths and weaknesses compared to earlier DMRG response methods. To understand these strengths and weaknesses better, 
we perform a series of benchmark
static and frequency-dependent polarisability calculations on 
oligo-diacetylenes that compare the behaviour of the earlier dynamical DMRG method with our
analytic response DMRG approach. Using our data we examine the scaling of the polarisability as a function of
the number of monomer units.

\section{Time-independent and time-dependent
density matrix renormalization group equations}

\begin{figure}
\includegraphics[width=3in]{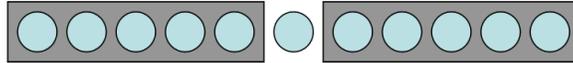}
\caption{One-site DMRG block configuration. $\mathbf{L}^n$ tensors
are associated with the left block, $\mathbf{R}^n$ tensors with
the right block, and the middle site is site $p$ in Eq. \ref{eq:ansatz}. \label{fig:dmrg_blocks}}
\end{figure}

The density matrix renormalisation group works with a variational ansatz
for the wavefunction $\Psi$. The simplest ansatz to analyse is the ``one-site'' form of the DMRG wavefunction \cite{white-onedot, zgid-twobody, Chan2002}.
For the block-configuration depicted in Fig. \ref{fig:dmrg_blocks}, the wavefunction takes the form
\begin{align}
|\Psi\rangle = \sum_{ \{ n \}} \mathbf{L}^{n_1} \ldots \mathbf{L}^{n_{p-1}}
\mathbf{C}^{n_p}
\mathbf{R}^{n_{p+1}} \ldots \mathbf{R}^{n_k} |n_1 \ldots n_k\rangle \label{eq:ansatz}
\end{align}
The $\mathbf{L}^n$ and $\mathbf{R}^n$ renormalisation tensors satisfy the orthogonality conditions
\begin{align}
\sum_n \mathbf{L}^{n\dag} \mathbf{L}^n &= \mathbf{1} \label{eq:l_ortho}\\
\sum_n \mathbf{R}^{n} \mathbf{R}^{n\dag} &= \mathbf{1} \label{eq:r_ortho}
\end{align}
and formally define the sequence of renormalisation transformations to obtain basis 
states $\{ l \}$, $\{ r\}$ for the left and right blocks in Fig. \ref{fig:dmrg_blocks}. (Note that in Eqs. (\ref{eq:l_ortho}), (\ref{eq:r_ortho}) we have dropped the sub-indices on $n$ as these conditions are not specific to any given site. We will use a similar convention
throughout to avoid a proliferation of unnecessary indices).
The coefficient tensor $\mathbf{C}^{n}$ gives  the expansion coefficients of the wavefunction in the superblock basis $\{l \} \otimes \{n_p\} \otimes \{r \}$. When
viewed as a flattened vector $\mathbf{c}$ it satisfies the normalisation condition $\mathbf{c}^\dag \mathbf{c}=1$. 

The DMRG energy is minimised when the tensors satisfy certain equations. For the coefficient vector,
this is a time-independent effective Schr\"odinger equation
\begin{equation}
\mathbf{H} \mathbf{c} = E \mathbf{c} \label{eq:eff_h}
\end{equation}
where the effective renormalised superblock Hamiltonian $\mathbf{H}$ satisfies $E = \langle \Psi | H |\Psi\rangle = \mathbf{c}^\dag \mathbf{H} \mathbf{c}$. The renormalisation tensors at each
position are defined from the coefficient tensor at the same position, i.e.  $\mathbf{C}^n$ defines $\mathbf{L}^n$ and 
$\mathbf{R}^n$, via  intermediate left and right density matrices. To obtain the left density matrix $\mathbf{D}_L$, 
we view the tensor $\mathbf{C}^n$ as a matrix $\mathbf{C}$ indexed by  $(ln), r$, where $l$ is the row index of $\mathbf{C}^n$, then
$\mathbf{D}_L = \mathbf{C} \mathbf{C}^\dag$. The right density matrix $\mathbf{D}_R$ is defined in a similar way, we view the 
tensor $\mathbf{C}^n$ as a matrix $\mathbf{C}$ indexed by $l, (nr)$, where $r$ is the column index of $\mathbf{C}^n$, and
$\mathbf{D}_R = \mathbf{C}^\dag \mathbf{C}$. The renormalisation tensors $\mathbf{L}^n$, $\mathbf{R}^n$, when viewed as matrices $\mathbf{L}, \mathbf{R}$ in the appropriate way, are obtained from the  $M$ eigenvectors (with largest
weights) of the the density matrix $\mathbf{D}_L$ and $\mathbf{D}_R$ respectively i.e.
\begin{align}
\mathbf{D}_L \mathbf{L} &= \mathbf{L} (\sigma_1 \ldots \sigma_M)_\text{diag}, \ \ \sigma_1 \geq \sigma_2 \ldots \geq \sigma_M, \ (\mathbf{L}_{(ln)i} = \mathbf{L}^n_{li}) \label{eq:l_tensor_eq} \\
\mathbf{D}_R \mathbf{R} &= \mathbf{R} (\sigma_1 \ldots \sigma_M)_\text{diag}, \ \ \sigma_1 \geq \sigma_2 \ldots \geq \sigma_M, \ (\mathbf{R}_{(rn)i} = \mathbf{R}^n_{ir}) \label{eq:r_tensor_eq} 
\end{align}
More explicitly, writing the eigenvectors of the left and right density matrices as $\mathbf{l}^i$, $\mathbf{r}^i$, 
\begin{align}
\mathbf{D}_L \mathbf{l}^i &= \mathbf{l}^i \sigma_i, \ \ \sigma_1 \geq \sigma_2 \geq \sigma_3 \ldots\\
\mathbf{D}_R \mathbf{r}^i &= \mathbf{r}^i \sigma_i, \ \ \sigma_1 \geq \sigma_2 \geq \sigma_3 \ldots
\end{align}
$\mathbf{L}^n$, $\mathbf{R}^n$ are  constructed by assigning the 
elements of the eigenvectors to the tensors in the following way
\begin{align}
 \mathbf{L}^n_{ji} &= \mathbf{l}^i_{(nj)}, \ \ i=1 \ldots M \label{eq:l_tensor} \\
 \mathbf{R}^n_{ij} &= \mathbf{r}^i_{(nj)}, \ \ i = 1 \ldots M \label{eq:r_tensor}
\end{align} 

In Ref. \cite{dmrg-lagrangian}, we showed that satisfying the
 solution conditions  Eqs. (\ref{eq:eff_h}), (\ref{eq:l_tensor_eq}), (\ref{eq:r_tensor_eq}) 
for $\mathbf{C}^n, \mathbf{L}^n, \mathbf{R}^n$ is formally equivalent to minimising the DMRG energy subject
to normalisation and the orthogonality constraints (\ref{eq:l_ortho}), (\ref{eq:r_ortho}). 
We can formally extend the DMRG theory to time-dependent scenarios by making stationary the 
Dirac-Frenkel action $\langle \Psi | i \partial/\partial t - H | \Psi\rangle$  \cite{mcweeny}
subject to the same normalisation and orthogonality constraints. (Interestingly, the Dirac-Frenkel
action has recently been independently rederived in the DMRG context in Ref. \cite{nishino-action}). For the coefficient vector
the time-evolution is then given by an effective time-dependent Schr\"odinger equation 
\begin{equation}
i \partial_t{\mathbf{c}} = \mathbf{H} \mathbf{c} \label{eq:eff_tdse}
\end{equation}
The corresponding $\mathbf{L}^n$ and $\mathbf{R}^n$ remain defined by Eqs. (\ref{eq:l_tensor_eq}), (\ref{eq:r_tensor_eq}).

\section{Coupled-perturbed density matrix renormalization group response equations}

We now consider the effect of an external perturbation. We start with
a time-independent perturbation $V$. In the superblock basis $\{l \} \otimes \{n_p\} \otimes \{r \}$, this yields
the effective perturbation $\mathbf{V}$ which satisfies $\langle \Psi | V | \Psi\rangle = \mathbf{c}^\dag \mathbf{V} \mathbf{c}$.

In response to this perturbation, the $\mathbf{L}^n, \mathbf{C}^n, \mathbf{R}^n$ tensors each can be expanded in
orders of $|V|$, giving
\begin{align}
\mathbf{L}^n &= \mathbf{L}^{n[0]} + \mathbf{L}^{n[1]} + \ldots\\
\mathbf{C}^n &= \mathbf{C}^{n[0]} + \mathbf{C}^{n[1]} + \ldots\\
\mathbf{R}^n &= \mathbf{R}^{n[0]} + \mathbf{R}^{n[1]} + \ldots
\end{align} 
Thus the  first-order DMRG
wavefunction for the block-configuration in Fig. \ref{fig:dmrg_blocks} takes the general form
\begin{align}
|\Psi^{[1]}\rangle &= \sum_{ \{ n\} }  \left[(\mathbf{L}^{n_1[1]}  \ldots \mathbf{C}^{n_p[0]} \ldots
\mathbf{R}^{n_k[0]}) + \ldots + (\mathbf{L}^{n_1[0]}  \ldots \mathbf{C}^{n_p[1]} \ldots \mathbf{R}^{n_k[0]}) \right. \nonumber \\
  &  \left. + \ldots + (\mathbf{L}^{n_1[0]}  \ldots \mathbf{C}^{n_p[0]} \ldots
\mathbf{R}^{n_k[1]})
\right] |n_1 n_2 \ldots n_k\rangle
\end{align}
We now derive the response equations satisfied by each of the quantities $\mathbf{L}^{n[1]}, \mathbf{C}^{n[1]}, \mathbf{R}^{n[1]}$. 
These are obtained by the perturbation expansion of the solution conditions 
(\ref{eq:eff_h}), (\ref{eq:l_tensor_eq}), (\ref{eq:r_tensor_eq}). For the
coefficient vector, this yields
\begin{align}
(\mathbf{H}^{[0]} + \mathbf{\Delta H}^{[1]} + \mathbf{V}^{[1]} + \ldots)(\mathbf{c}^{[0]} +
\mathbf{c}^{[1]} + \ldots) = (E^{[0]} + E^{[1]} +
\ldots)(\mathbf{c}^{[0]} +
\mathbf{c}^{[1]} + \ldots)
\end{align}
Note the first-order change in the Hamiltonian $\mathbf{\Delta H}^{[1]}$. This arises because the
effective Hamiltonian in the superblock basis $\mathbf{H}$ depends on the renormalisation tensors 
$\mathbf{L}^{n}, \mathbf{R}^n$ (which define the renormalised basis) and so first-order changes in those tensors lead to a first-order change in the effective Hamiltonian. (The  construction of $\mathbf{\Delta H}^{[1]}$ is described later in Sec. \ref{sec:algo}). 
Gathering first-order terms and enforcing intermediate normalisation
through the projector $\mathbf{Q} = \mathbf{1} - \mathbf{c}^{[0]} \mathbf{c}^{[0]\dag}$ gives
\begin{align}
(\mathbf{H}^{[0]} - E^{[0]} \mathbf{1}) \mathbf{c}^{[1]}
= -\mathbf{Q} (\mathbf{\Delta H}^{[1]} + \mathbf{V}^{[1]})
\mathbf{c}^{[0]} \label{eq:response}
\end{align}
Because $\mathbf{\Delta H}^{[1]}$ depends on the first-order wavefunction through its dependence on
the $\mathbf{L}^n, \mathbf{R}^n$ tensors, Eq. (\ref{eq:response}) must be solved self-consistently. It
is therefore a coupled-perturbed response equation, analogous to the coupled-perturbed orbital equations
 that arise in the Hartree-Fock theory of response.

The first-order coefficients $\mathbf{C}^{n[1]}$ define first-order renormalisation tensors at the same site
 $\mathbf{L}^{n[1]}, \mathbf{R}^{n[1]}$. Viewing $\mathbf{C}^{n[0]}, \mathbf{C}^{n[1]}$
as a matrices in the appropriate fashion, we obtain first-order left and right density matrices
\begin{align}
\mathbf{D}_L^{[1]} & = \mathbf{C}^{[0]} \mathbf{C}^{[1]\dag}+\mathbf{C}^{[1]} \mathbf{C}^{[0]\dag}, 
\ (\mathbf{C}_{(nl),r} = \mathbf{C}^n_{lr}) \label{eq:left_dm1}\\
\mathbf{D}_R^{[1]} & = \mathbf{C}^{[0]^\dag} \mathbf{C}^{[1]}+\mathbf{C}^{[1]\dag} \mathbf{C}^{[0]}, 
\ (\mathbf{C}_{l,(nr)} = \mathbf{C}^n_{lr}) \label{eq:right_dm1}
\end{align}
In response to the change in the density matrices, the eigenvectors have a perturbation expansion
\begin{align}
\mathbf{l}^i &= \mathbf{l}^{i[0]} + \mathbf{l}^{i[1]} + \ldots \\
\mathbf{r}^i &= \mathbf{r}^{i[0]} + \mathbf{r}^{i[1]} + \ldots
\end{align}
and we can set up corresponding response equations
\begin{align}
(\mathbf{D}_L^{[0]}- \sigma_i \mathbf{1} ) \mathbf{l}^{i[1]} &= -\mathbf{Q}_L 
\mathbf{D}_L^{[1]} \mathbf{l}^{i[0]} \label{eq:dm_response1} \\
(\mathbf{D}_R^{[0]}- \sigma_i \mathbf{1} ) \mathbf{r}^{i[1]} &= -\mathbf{Q}_R 
\mathbf{D}_R^{[1]} \mathbf{r}^{i[0]} \label{eq:dm_response2}
\end{align}
where $\mathbf{Q}_L, \mathbf{Q}_R$ project out the span of $\mathbf{D}_L, \mathbf{D}_R$ respectively, i.e.
 $\mathbf{Q}_L =\mathbf{1} - \sum_{i=1}^M \mathbf{l}^{i[0]} \mathbf{l}^{i[0]\dag}, 
\mathbf{Q}_R =\mathbf{1} - \sum_{i=1}^M \mathbf{r}^{i[0]} \mathbf{r}^{i[0]\dag}$.
We  assign the elements  of each of the $M$ perturbed vectors $\mathbf{l}^{i[1]}, \mathbf{r}^{i[1]}$
according to Eq. (\ref{eq:l_tensor}), (\ref{eq:r_tensor}),
to define $\mathbf{L}^{n[1]}, \mathbf{R}^{n[1]}$.
The response equations for a time-dependent perturbation may be
obtained in an analogous way as above. We consider for simplicity a perturbation with
a single Fourier component,
\begin{equation}
V(t) = Ve^{i\omega t} + V^* e^{-i\omega t}
\end{equation}
We  expand 
the $\mathbf{L}^n, \mathbf{C}^n, \mathbf{R}^n$ tensors in terms of orders of $|V|$, 
\begin{align}
\mathbf{L}^{n}(t) &= (\mathbf{L}^{n[0]} + \mathbf{L}^{n[1]}(t) + \ldots)
e^{-i E^{[0]} t} \\
\mathbf{C}^{n}(t) &= (\mathbf{C}^{n[0]} + \mathbf{C}^{n[1]}(t) + \ldots)
e^{-i E^{[0]} t} \\
\mathbf{R}^{n}(t) &= (\mathbf{R}^{n[0]} + \mathbf{R}^{n[1]}(t) + \ldots)
e^{-i E^{[0]} t} 
\end{align}
For the coefficient vector, we substitute this expansion into the effective
time-dependent Schr\"odinger equation (\ref{eq:eff_tdse}) and identify terms
with frequencies $\omega$, $-\omega$, giving
\begin{align}
(\mathbf{H}^{[0]} - (E^{[0]} + \omega) \mathbf{1}) \mathbf{c}^{[1]}(\omega)
&= -\mathbf{Q} (\mathbf{\Delta H}^{[1]}(\omega) + \mathbf{V}^{[1]} )  \mathbf{c}^{[0]}\label{eq:time_response1}\\
(\mathbf{H}^{[0]} - (E^{[0]} -\omega) \mathbf{1}) \mathbf{c}^{[1]}(-\omega)
&= -\mathbf{Q} (\mathbf{\Delta H}^{[1]}(-\omega) + \mathbf{V}^{[1]*})
\mathbf{c}^{[0]} \label{eq:time_response2}
\end{align}
where $\mathbf{Q}$ is the projector defined in Eq. (\ref{eq:response}). The first-order frequency perturbed wavefunctions
then define first-order perturbed density matrices $\mathbf{D}_L(\omega), \mathbf{D}_L(-\omega), \mathbf{D}_R(\omega), \mathbf{D}_R(-\omega)$, which can be used to obtain $\mathbf{L}^{n[1]}(\omega), \mathbf{L}^{n[1]}(-\omega),
\mathbf{R}^{n[1]}(\omega), \mathbf{R}^{n[1]}(-\omega)$ through Eqs. (\ref{eq:dm_response1}), (\ref{eq:dm_response2}).

\subsection{Response properties}
\label{sec:properties}

Once we obtain the first-order response of the DMRG wavefunction we can evaluate response properties of interest. We
take as our example here the dipole-dipole response function or polarisability. For a uniform static electric field $\mathcal{E}_i$,
the dipole moment is expanded as
\begin{align}
\mu_i = \mu_i^{[0]} + \sum_{j} \alpha_{ij} \mathcal{E}_j + \ldots, \ \ i,j \ldots \in x, y, z
\end{align}
which defines the static polarisability $\alpha_{ij}$ as the first-order change in the dipole moment. Within
the DMRG response theory, the polarisability is therefore obtained as
\begin{align}
\alpha_{ij} = \mathbf{c}^{[0]\dag} \mathbf{\mu}_i^{[0]} \mathbf{c}^{[1]}_j + 
\mathbf{c}^{[1]\dag}_j \mathbf{\mu}_i^{[0]} \mathbf{c}^{[0]} + \mathbf{c}^{[0]\dag} \mathbf{\mu}_{i(j)}^{[1]} \mathbf{c}^{[0]} \label{eq:polar}
\end{align}
Here $\mathbf{\mu}_i$ is the effective dipole operator in the superblock basis, and $\mathbf{c}^{[1]}_j$ is the first-order
wavefunction in response to an electric field in the $j$ direction. Note the additional contribution
$\mathbf{\mu}_{i(j)}^{[1]}$. This is the change in the effective dipole operator $\mathbf{\mu}_i$ due to the
response of the $\mathbf{L}^n$, $\mathbf{R}^n$ tensors to an applied field in the $j$ direction. This quantity is constructed
in a similar way to the effective Hamiltonian $\mathbf{\Delta H}^{[1]}$.

For a frequency dependent electric field $\mathcal{E}_i (t)$, we expand the dipole moment as
\begin{align}
\mu_i(t) = \mu_i^{[0]} + \sum_j \int \ d\omega e^{-i \omega t} \alpha_{ij} (\omega) \mathcal{E}_j(\omega)  + \ldots \ \ i,j \ldots \in x, y, z
\end{align}
where $\alpha_{ij}(\omega)$ and $\mathcal{E}_j(\omega)$ are the $\omega$ frequency components
of the frequency dependent polarisability and electric field. $\alpha_{ij}(\omega)$ contains two contributions, one from
the $e^{i\omega t}$ component of the applied perturbation, one from the $e^{-i\omega t}$ component. The final expression for
$\alpha_{ij}(\omega)$ therefore reads as
\begin{align}
\alpha_{ij}(\omega) &= G_{ij}(\omega) + G_{ij}(-\omega)\\
G_{ij}(\omega) &= \mathbf{c}^{[0]\dag} \mathbf{\mu}_i^{[0]} \mathbf{c}^{[1]}_j(\omega) + 
\mathbf{c}^{[1]\dag}_j(\omega) \mathbf{\mu}_i^{[0]} \mathbf{c}^{[0]} + \mathbf{c}^{[0]\dag} \mathbf{\mu}_{i(j)}^{[1]}(\omega) \mathbf{c}^{[0]}
\end{align}
$G_{ij}(\omega)$ and $G_{ij}(\omega)$ are obtained from two separate response calculations, solving Eq. (\ref{eq:time_response1}), (\ref{eq:time_response2}) respectively.

\subsection{Comparison to other  DMRG response theories}
\label{sec:comparison}
So far we have derived a DMRG theory of response that was based on  expanding the solution
conditions satisfied by the DMRG wavefunction in terms of the applied perturbation. This corresponds to
an \textit{analytic} theory of response in the following way. Consider a
 time-independent perturbation for simplicity.
Let us consider minimising the energy of the DMRG wavefunction, for some fixed number of states $M$, 
with respect to the full Hamiltonian (with the perturbation) $H = H^{[0]} + \lambda V^{[1]}$ where
$\lambda$ is used to scale the strength of the perturbation. This gives a wavefunction $\Psi(\lambda)$ and
an energy $E(\lambda)$. The first-order wavefunction $\Psi^{[1]}$, and corresponding first-, second-, and third-order
energies calculated with the analytic DMRG response theory correspond exactly to the following derivatives
\begin{align}
\Psi^{[1]} &= \left.\frac{\partial \Psi(\lambda)}{\partial \lambda}\right|_{\lambda=0}\\
E^{[1]} & = \left.\frac{\partial E(\lambda)}{\partial \lambda}\right|_{\lambda=0}\\
E^{[2]} & = \left.\frac{\partial^2 E(\lambda)}{\partial \lambda^2}\right|_{\lambda=0} \\
E^{[3]} & = \left.\frac{\partial^3 E(\lambda)}{\partial \lambda^3}\right|_{\lambda=0}
\end{align}
Analogous statements for  time-dependent perturbations
can be made by considering an appropriate quasi-energy \cite{rice1991cfd, sasagane1993hor,olsen-review}.

The analytic approach to DMRG response  does not represent the only way to obtain response within the DMRG. 
Existing  DMRG response methods use various related \textit{adaptive basis} approaches,  commonly  in two categories,  
 the Lanczos vector method \cite{hallberg1995dma}, and the  dynamical density matrix renormalisation group \cite{jeckelmann2002ddm}. 
The dynamical density matrix renormalisation group is 
 established as the most accurate approach to response properties and   we shall focus on it here. (Note
the dynamical density matrix renormalisation group and correction vector methods 
\cite{ramasesha1996sdm, kuhner1999dcf, jeckelmann-review} are essentially the same but differ
in the algorithm used to solve the response equations. In fact,
if the response quantities are evaluated using a quadratic functional
of the correction vector such as Eq. (\ref{eq:quad}), it is possible to obtain
quadratic errors with the correction vector method without the explicit minimisation as used in the dynamical DMRG).


In the dynamical DMRG the ansatz for the zeroth and first-order wavefunction
are both modified relative to the unperturbed  DMRG wavefunction, i.e.
\begin{align}
 |\Psi^{[0]}\rangle &= \sum_{\{n\}}  \tilde{\mathbf{L}}^{n_1[0]} 
 \ldots \tilde{\mathbf{C}}^{n_p [0]} \ldots  \tilde{\mathbf{R}}^{n_k[0]}|n_1 n_2 \ldots n_k\rangle \label{eq:ddmrg_wf0} \\
 |\Psi^{[1]}\rangle &= \sum_{\{n\}}  \tilde{\mathbf{L}}^{n_1[0]}
 \ldots \tilde{\mathbf{C}}^{n_p [1]} \ldots \tilde{\mathbf{R}}^{n_k[0]}|n_1 n_2 \ldots n_k\rangle \label{eq:ddmrg_wf1}
\end{align}
The tildes indicate that the $\tilde{\mathbf{L}}^n$, $\tilde{\mathbf{C}}^n$, $\tilde{\mathbf{R}}^n$ 
tensors appearing in Eqs. (\ref{eq:ddmrg_wf0}), even
for the zeroth order wavefunction, do not correspond to the same tensors obtained in a DMRG calculation without the perturbation. 
The zeroth and first-order coefficient vectors are obtained from the effective Schr\"odinger equation (\ref{eq:eff_h}) and an uncoupled
response equation,  e.g.
\begin{align}
(\mathbf{H}^{[0]} - (E^{[0]} + \omega) \mathbf{1}) \mathbf{c}^{[1]}(\omega)
&= -\mathbf{Q} \mathbf{V}^{[1]}  \mathbf{c}^{[0]}\label{eq:ddmrg_response}
\end{align}
The dynamical DMRG ansatz is able to capture the response of the $\mathbf{L}^n$ and $\mathbf{R}^n$ tensors in an average way, because
it uses  $\tilde{\mathbf{L}}^n$, $\tilde{\mathbf{R}}^n$ that are different from those in the unperturbed DMRG calculation. 
Specifically, the left and right renormalisation tensors  at each block configuration are obtained
as eigenvectors of \textit{modified} left and right density matrices, where  the density matrices corresponding to
$\mathbf{c}^{[0]}, \mathbf{c}^{[1]}, \mathbf{v}=\mathbf{V}^{[1]}\mathbf{c}^{[0]}$ are all averaged together i.e. for $\mathbf{D}_L$
\begin{align}
\mathbf{D}_L = \alpha \mathbf{C}^{[0]} \mathbf{C}^{[0]\dag}  + \beta \mathbf {C}^{[1]} \mathbf{C}^{[1]\dag} +
\gamma (\mathbf{V} \mathbf{c}^{[0]}) (\mathbf{V} \mathbf{c}^{[0]})^\dag
\end{align}
where $\alpha+\beta+\gamma=1$ and in the last term we are interpreting the perturbation multiplied by the zeroth order wavefunction $\mathbf{V}^{[1]}\mathbf{c}^{[0]}$
as a matrix in the same way as $\mathbf{c}^{[0]}$ is interpreted as a matrix. (Note that the above is for real frequencies; when
for complex frequencies, one typically separates the imaginary and real contributions of the response vector \cite{jeckelmann2002ddm}). Because the
density matrix contains information on the perturbation and the response, 
 the DMRG basis is ``adapted'' to the
perturbation being considered. While this is very simple to implement within a standard DMRG algorithm and has proven
very successful, one drawback relative
to the analytic response approach is that a single
set of DMRG basis states is being used to represent several quantities, including both the zeroth order and response vectors. 
For this reason, we can expect some loss of accuracy 
with this method for small $M$ calculations relative to the analytic response method.

\section{Implementation}
\label{sec:algo}

We have implemented the analytic DMRG response theory as described above. This consists of three 
parts:  solving the coupled-perturbed equation (\ref{eq:response}) for the first-order coefficient vector $\mathbf{c}^{[1]}$, 
 solving for the first-order renormalisation tensors $\mathbf{L}^{n[1]}$, $\mathbf{R}^{n[1]}$ (\ref{eq:l_tensor_eq}), (\ref{eq:r_tensor_eq}), and constructing the first-order effective Hamiltonian $\mathbf{\Delta H}^{[1]}$ and necessary intermediates, as well as
other first-order operators needed for properties (e.g. $\mathbf{\mu}_{i(j)}^{[1]}$ in Eq. \ref{eq:polar}). The first two parts are quite straightforward: we solve the coupled-perturbed equation (\ref{eq:response}) using a Krylov subspace
iterative solver with preconditioning, and to obtain the first-order renormalisation 
tensors (\ref{eq:dm_response1}), (\ref{eq:dm_response2})
we use  explicit Rayleigh-Schr\"odinger expressions for the first-order density matrix eigenvectors
\begin{align}
\mathbf{l}^{i[1]} &= \sum_{j={M+1}} - \frac{\mathbf{l}^{j[0]^\dag} \mathbf{D}_L^{[1]} \mathbf{l}^{i[0]}}{\sigma_j^{[0]} - \sigma_i^{[0]}} \mathbf{l}^{j[0]} \\
\mathbf{r}^{i[1]} &= \sum_{j={M+1}} - \frac{\mathbf{r}^{j[0]^\dag} \mathbf{D}_R^{[1]} \mathbf{r}^{i[0]}}{\sigma_j^{[0]} - \sigma_i^{[0]}} \mathbf{r}^{j[0]}
\end{align}
We now focus on the implementation to obtain $\mathbf{\Delta H}^{[1]}$ and related quantities such as $\mathbf{\mu}_{i(j)}^{[1]}$. 
We recall that the effective Hamiltonian $\mathbf{H}^{[0]}$
is expressed as a tensor product of operators on the left and right blocks (we
 consider the single-site $\bullet$ in the block configuration Fig. \ref{fig:dmrg_blocks} 
to be part of the left block for simplicity) 
\begin{align}
\mathbf{H} = \sum_{ij} w_{ij}  \mathbf{O}_L^i \otimes \mathbf{O}_R^j
\end{align}
where $\mathbf{O}_L$ acts only the left block and $\mathbf{O}_R$ acts only on the right block, and we assume that
$\otimes$ takes into account the appropriate parity factors associated with the fermion character of the operators (see e.g.
Ref. \cite{SCHOLLWOCK:2005:_dmrg, Chan2002}).  The first-order
Hamiltonian is constructed from the response of the operators $\mathbf{O}_L, \mathbf{O}_R$, through
\begin{align}
\mathbf{\Delta H}^{[1]} = \sum_{ij} w_{ij} (\mathbf{O}_L^{i[0]} \mathbf{O}_R^{j[1]} + \mathbf{O}_L^{i[1]} \mathbf{O}_R^{j[0]})
\end{align}
We therefore need to calculate the first-order  operators $\mathbf{O}_L^{[1]}$, $\mathbf{O}_R^{[1]}$. 
These are built up sequentially through the blocking steps in the sweep much like the zeroth order operators. The
renormalisation transformation $\mathcal{R}$ of the first-order operator at a given block configuration in a left$\to$right sweep, 
is given by
\begin{align}
\mathcal{R}[\underline{\mathbf{O}}_L^{[1]}] = \mathbf{L}^{[0]\dag} \underline{\mathbf{O}}_L^{[1]} \mathbf{L}^{[0]} + \mathbf{L}^{[1]\dag} 
\underline{\mathbf{O}}_L^{[0]} \mathbf{L}^{[0]}
+ \mathbf{L}^{[0]\dag} \underline{\mathbf{O}}_L^{[0]} \mathbf{L}^{[1]} \label{eq:transform_1}
\end{align}
where we have used the underline to indicate that the operators refer to blocked operators (i.e. for the left block plus the single-site), and the renormalisation
tensors are interpreted as matrices $\mathbf{L}$ as described in Eq. (\ref{eq:l_tensor_eq}). At the beginning of the left$\to$right sweep,
$\mathbf{O}_L^{[1]}=0$ for all such operators.
Analogous expressions hold for the right$\to$left sweep and the operators $\mathbf{O}_R$.

The full sweep algorithm for the DMRG analytic response can be summarised as follows:
\begin{enumerate}
\item Converge a standard DMRG algorithm for the state of interest and store all intermediate zeroth-order operators $\mathbf{O}_L^{[0]}$,
$\mathbf{O}_R^{[0]}$ and tensors $\mathbf{L}^{n[0]}$, $\mathbf{C}^{n[0]}$, $\mathbf{R}^{n[0]}$.
\item Set all $\mathbf{O}_L^{[1]}, \mathbf{O}_R^{[1]}$ = 0
\item Start a sweep (left$\to$right)
\begin{itemize}
\item Set all $\mathbf{O}_L^{[1]}$ to 0
\item At each block configuration:
\item Solve coupled perturbed response equation, Eq. (\ref{eq:response}). $\mathbf{\Delta H}^{[1]}$ is constructed
using current best guesses for $\mathbf{O}_L^{[1]}$, $\mathbf{O}_R^{[1]}$
\item Solve for perturbed density matrix eigenvectors and $\mathbf{L}^{n[1]}$, Eq. (\ref{eq:dm_response1})
\item Update all $\mathbf{O}_L^{[1]}$ using Eq. (\ref{eq:transform_1})
\end{itemize}
\item Start a sweep (right$\to$left), analogous to (left$\to$right) sweep
\item Loop to 3. until convergence.
\item Evaluate response properties (e.g. as in Sec. \ref{sec:properties})
\end{enumerate}
We note that the cost of a single sweep for the analytic response has the same
order of computational and storage cost as an ordinary sweep in the DMRG
calculation, which, for the \textit{ab-initio} Hamiltonian is $O(M^3 k^3) +O(M^2 k^4)$ computation, $O(M^2 k^2)$ memory, and $O(M^2 k^3)$ disk, where $k$ is the number of correlated orbitals.
The memory cost is roughly twice that for the calculation of the energy because of 
storage of the first-order operators as well as the zeroth-order  operators.

\section{Static and frequency-dependent polarizabilities of oligo-di-acetylenes}

As an initial test of the analytic DMRG response theory and implementation, we have calculated static and
frequency-dependent longitudinal polarisabilities of several oligo-di-acetylenes using the analytic 
DMRG response theory, the dynamical DMRG method, and the linear-response coupled cluster method.
 Long oligo-di-acetylenes are  of
interest due to their large third-order non-linear polarisability \cite{bredas-review}. While we will calculate only the linear
polarisability here, the same analytic derivative techniques can in principle be extended to higher order polarisabilities
and non-linear optical response.

\begin{figure}
\includegraphics[width=5in]{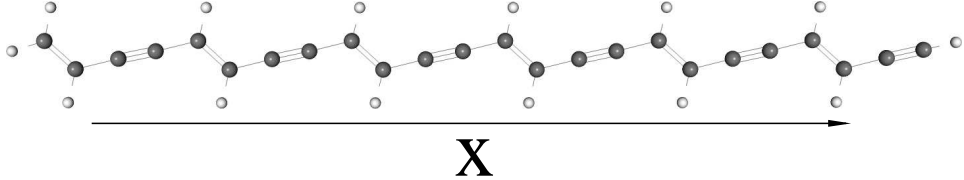}
\caption{Oligo-di-acetylenes, with the long-axis moment of inertia  aligned with the $x$-coordinate. This
is the axis along which the
polarisabilities are evaluated. \label{fig:pda}}
\end{figure}

We carried out calculations on short all-trans oligo-di-acetylenes (ODAs), 2-ODA $\mathrm{C_8H_6}$, 4-ODA $\mathrm{C_{16}H_{10}}$, 6-ODA $\mathrm{C_{24}H_{14}}$.
Optimised geometries were obtained at
the density functional theory B3LYP \cite{b3,lyp} level in a correlation consistent
Dunning double-zeta (cc-pVDZ) basis \cite{emsl}.
Subsequent Hartree-Fock, DMRG, and coupled cluster (CC) calculations were carried out in a minimal STO-6G Gaussian basis \cite{sto,emsl}.
We realise that this basis is too small for the quantitative calculation of polarisabilities, but it has been chosen to
enable a preliminary study. Also, we note that  qualitative trends in polarisabilities can be  captured using rather small basis sets of split-valence
quality \cite{bredas-review}.
The Hartree-Fock calculations were used
 to
determine molecular orbitals with $\sigma$ and $\pi$ character. 
All $\sigma$ orbitals were kept frozen in the DMRG response calculations,
and the $\pi$ orbitals were localised.
 Calculated polarisabilities refer to the $\alpha_{xx}$ component, where the $x$-axis is aligned with the long
moment of inertia axis  of the molecules (see Fig. \ref{fig:pda}).
The DMRG response calculations used an active space of $p_z$ orbitals only, corresponding to an (8e, 8orb) active space for 2-ODA, a (16e, 16orb) active space
for 4-ODA, and a (24e, 24orb) active space for 6-ODA. 
For the analytic response DMRG calculations using $M$ states we first converged a ground-state DMRG calculation with $M$ states using the one-site
algorithm, and used this as the starting point for the response calculation. 

In addition to the analytic response DMRG calculations, we  carried out   calculations
using the dynamical DMRG method for comparison. The dynamical DMRG polarisabilities were obtained by solving the linear response equation
in the dynamical DMRG basis $(\omega \mathbf{1} - \mathbf{H}^{[0]} )\mathbf{c}^{[1]})_i = \mathbf{Q} \mathbf{\mu}_i \mathbf{c}^{[0]}$ just 
as in the correction vector method, but the resulting polarisabilities were evaluated using the quadratic functional 
\begin{align}
G_{ij} = \mathbf{c}^{[1]\dag}_i (\omega \mathbf{1} - \mathbf{H}^{[0]}) \mathbf{c}^{[1]}_j +
\mathbf{c}^{[0]\dag} \mathbf{\mu}_i  \mathbf{c}^{[1]}_j + \mathbf{c}^{[1]\dag}_j \mathbf{\mu}_i  \mathbf{c}^{[0]}j \label{eq:quad}
\end{align}
which ensures that the obtained polarisability is quadratic in the error in $\mathbf{c}^{[1]}$ \cite{koch1991acf, sellers}, 
which is the hallmark of the dynamical DMRG  approach.
For comparison, we also computed linear-response restricted coupled cluster polarisabilities at the singles and doubles level \cite{koch1990eec},
both at the all electron level, and within the $p_z$ active space only, using
the \textsc{Psi3} \cite{PSI3} package.

We note one issue that arises with the response DMRG calculations in our initial implementation as opposed to ordinary ground-state DMRG calculations.
 In ground-state DMRG calculations with the one-site algorithm, we are generally able to converge the DMRG energy from sweep to sweep
to very high accuracy, e.g. nanoHartrees. However, in our initial response implementation, we were not able to 
converge the calculated polarisabilities to similar accuracy. 
Typically the forward and backwards sweeps would converge to 
somewhat different results, and even between consecutive forwards (or backwards) sweeps, the polarisability would oscillate somewhat.
This was true both for the dynamical DMRG and the analytic response DMRG calculations.
The oscillation can be quite severe, particularly for small $M$ calculations and for higher frequencies that
are nearer to  a pole (e.g. at frequency $\omega=0.2$ a.u.) and reflects the greater sensitivity of the response calculation to the discarded states in the density matrix.
In our results, we
report the average polarisability of the last 4 sweeps, together with twice the standard deviation. These
results are reported in table \ref{tab:results}.



\begin{table}
\caption{Static and frequency dependent polarisabilities in a.u. of oligo-di-acetylenes, with 2, 4, 6 monomers (2-ODA, 4-ODA, 6-ODA).
$D$ stands for dynamical DMRG, $A$ stands for analytic response theory. $\omega$ is the frequency (in a.u.), $M$ refers
to the number of states in the DMRG wavefunction. \textbf{The numbers in brackets do not represent intrinsic truncation error
 from finite $M$} but
represent the numerical convergence of the DMRG sweep, since
the forwards and backwards sweeps typically
converge to slightly different results. The bracketed number
is twice the standard deviation ($2\sigma$) in
the last 4 forward and backwards sweeps. See text for further discussion.
\label{tab:results}}
\begin{center}
\begin{tabular*}{5.8in}{@{\extracolsep{\fill}} l l| c c| c c| c c }
\hline
\hline
    &       &\multicolumn{2}{c}{2-ODA}&\multicolumn{2}{c}{4-ODA} & \multicolumn{2}{c}{6-ODA} \\
\hline
$\omega$ & M     & D       & A  & D $(2\sigma)$      & A $(2\sigma)$ & D $(2\sigma)$ & A $(2\sigma)$ \\
\hline
0.00& 25    & 52.77  & 52.89  & 144.16 (0.03) & 145.21 (0.04)  & 354.28 (17.96)& 243.65 (0.06) \\
    & 50    & 52.89  & 52.89  & 146.07 (0.01) & 145.74 (0.09)  & 246.04 (0.02) & 245.06 (0.07) \\
    & 250   & 52.88  & 52.88  & 145.75 (0.01) & 145.80 (0.01)  & 245.20 (0.00) & 245.27 (0.03) \\
    & 1000  & n.a.   & n.a.   & 145.77 (0.01) & 145.81 (0.00)  & 245.13 (0.10) & 245.14 (0.02) \\
    & LR-CCSD  &  \multicolumn{2}{c|}{53.38}&   \multicolumn{2}{c|}{148.15} & \multicolumn{2}{c}{249.67}  \\
\hline
0.05& 25    & 53.98  & 53.96  & 148.46 (0.02) & 149.80 (0.04) & 449.82 (35.15) & 252.00 (0.14) \\
    & 50    & 54.07  & 54.07  & 150.64 (0.01) & 150.26 (0.07) & 254.61 (0.02)  & 253.62 (0.13) \\
    & 250   & 54.06  & 54.07  & 150.37 (0.00) & 150.39 (0.04) & 253.87 (0.00)  & 253.92 (0.02)\\
    & LR-CCSD  & \multicolumn{2}{c|}{54.62}& \multicolumn{2}{c|}{153.19}  & \multicolumn{2}{c}{259.40} \\
\hline
0.10& 25    & 57.83  & 57.57  & 163.62 (0.03) & 165.42 (0.13) & 462.00 (22.55) & 282.05 (0.25) \\
    & 50    & 57.99  & 57.99  & 166.46 (0.02) & 166.11 (0.05) & 284.81 (0.03)  & 283.96 (0.22) \\
    & 250   & 57.99  & 58.00  & 166.19 (0.00) & 166.23 (0.02) & 284.30 (0.00)  & 284.26 (0.21) \\
    & LR-CCSD  & \multicolumn{2}{c|}{58.72}&   \multicolumn{2}{c|}{170.76}     &  \multicolumn{2}{c}{294.16}\\
\hline
0.15& 25    & 65.85  & 64.97  & 195.14 (0.07) & 201.06 (0.17) & 557.18 (114.72) & 353.66 (0.57) \\
    & 50    & 66.07  & 66.06  & 202.51 (0.03) & 202.24 (0.09) & 357.02 (0.05)   & 356.37 (0.20) \\
    & 250   & 66.05  & 66.08  & 202.45 (0.00) & 202.49 (0.04) & 357.26 (0.00)   & 357.10 (0.10)\\
    & LR-CCSD  & \multicolumn{2}{c|}{67.22}   & \multicolumn{2}{c|}{212.20}    & \multicolumn{2}{c}{381.68} \\
\hline
0.20& 25    & 82.03  & 79.89  & 279.06 (0.35) & 294.06 (0.89) & 520.61 (84.68) & 564.50 (1.38) \\
    & 50    & 82.57  & 82.54  & 296.86 (0.62) & 295.83 (1.67) & 564.25 (16.84) & 566.94 (0.89) \\
    & 250   & 82.56  & 82.60  & 296.71 (0.55) & 296.44 (0.06) & 571.44 (0.71)  & 571.63 (1.73) \\
    & LR-CCSD  & \multicolumn{2}{c|}{84.83}   & \multicolumn{2}{c|}{328.71}        & \multicolumn{2}{c}{682.10}  \\
\hline
\hline
\end{tabular*}
\end{center}
\end{table}

From table \ref{tab:results} we  make the following observations about the relative performance of the analytic DMRG
response method relative to the dynamical DMRG method that has been commonly used. 
For small $M$ (e.g. $M$=25) the analytic DMRG response method is clearly superior. Whereas the dynamical DMRG method produces  poor polarisabilities for $M$=25, in error by more
than 50\% in some cases,
the analytic DMRG polarisabilities are quite reasonable at $M$=25 and typically in error by less than 1\%. This is consistent with our discussion in section \ref{sec:comparison} where we argue the the dynamical DMRG method suffers from  using the same set of DMRG 
basis states to represent
both the zeroth order DMRG vector as well as the response and perturbation vectors. Thus, for small $M$ there simply are not enough DMRG
states to yield a meaningful result in the dynamical DMRG. Both methods converge  as $M$ increases. 
For the most accurate calculations ($M$=250), although both methods perform well,  the
dynamical DMRG polarisabilities appear slightly better than the analytic  DMRG  polarisabilities. However,
 this  appears to be related
to the   instabilities in the  convergence of  the analytic DMRG response sweeps; whereas the
oscillations in the dynamical DMRG sweeps vanish for larger $M$, they still remain 
 for the analytic DMRG sweeps. From the $2\sigma$ values, we see that currently we
can only  conclude  that the analytic and dynamical DMRG response methods are comparable for larger $M$.

Observing the trends in the polarisabilities, we see that the polarisabilities increase as the applied frequency increases which
is what one would expect since we are approaching the 
first excitonic ${}^1B_u$ pole. We are not able to converge our response calculations very close to a pole because
of the large norm in $\mathbf{c}^{[1]}$. 
 The standard solution to this is to include a small imaginary broadening in $\omega$. However, a straightforward incorporation of
 broadening leads to complex operators in the analytic theory which we have not yet implemented. 

It is often the case that one wishes to determine an entire spectrum, i.e. some response property for a very large range of $\omega$. While in the dynamical DMRG this is usually performed by scanning through $\omega$ (with some small imaginary component) and performing
a response calculation for each frequency, it may be more appropriate in the analytic response approach to adopt a different strategy. 
The coupled-perturbed response equations may be viewed as a   linear eigenvalue problem
for the excitation energies (i.e. poles) and may
be solved in this way, in the same way that the time-dependent Hartree-Fock or time-dependent density functional
 equations are solved as an eigenvalue problem
to obtain  excitation energies. Once a sufficient number of poles are obtained, the spectrum can then be reconstructed analytically.

Comparing the DMRG polarisabilities and the coupled cluster polarisabilities, we see that the coupled cluster polarisabilities
are generally quite good even at the singles and doubles level. (They appear to consistently overestimate the polarisability by
only a  few percent).
This is not  surprising since by virtue of the one-electron
nature of the dipole operator, the linear polarisability only samples states with single-excitation character relative to the
ground-state. Such excited states are well captured by CCSD theory. However, earlier studies indicate that the overall spectrum  in
conjugated systems (including
e.g. doubly excited and triplet excited states) is poorly reproduced by coupled cluster theory \cite{chan-jon}, and so we would expect much
larger discrepancies between the CC and DMRG description of  third-order non-linear optical response. 

\begin{figure}
\includegraphics[width=3in]{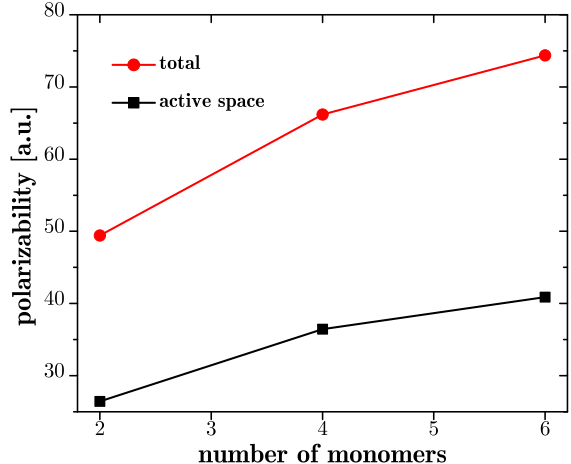}
\caption{Scaling of total and active space polarisabilities per monomer. \label{fig:scaling}}
\end{figure}

In Fig. \ref{fig:scaling} we plot the static active space and total polarisabilities ($\omega=0$) per monomer 
calculated using the analytic DMRG response theory
 as a function of the number of di-acetylene 
monomers in the calculation. The total polarisability for the DMRG calculations is obtained using the core-correction from the linear-response coupled cluster calculations i.e.
\begin{align}
\alpha^{\text{tot}}_{\text{DMRG}} = \alpha^{\text{tot}}_{CC} - \alpha^{\text{act}}_{CC} + \alpha^{\text{act}}_{DMRG}
\end{align}
We see a slow saturation of the polarisability per monomer as a function of the chain length, although the polarisability is
not yet  fully saturated at the 6-ODA level. While larger basis set calculations and calculations on longer chains
are necessary to obtain a definitive conclusion, we note that 
our results are consistent with early semi-empirical calculations
which indicate an onset of saturation between 2-ODA and 3-ODA \cite{kirtman-ijqc}.

\section{Conclusions}

In the current work we have described an analytic approach to the calculation of response quantities in the density matrix
renormalisation group. The analytic response method is familiar from other electronic structure theories
 but has not  so far been developed
within the density matrix renormalization group. The analytic response implementation 
does not change the computational cost of the ground-state DMRG
calculation by more than a constant factor. Compared to the popular dynamical density matrix renormalisation group  approach
we find that the analytic response method produces considerably more accurate response quantities when using a small number of DMRG
states, without any greater computational cost. While 
it is simpler within the dynamical DMRG  to implement  higher-order response properties and complex frequencies, based on our
investigations, the improved accuracy of the analytic response approach may 
justify the additional implementation effort.
 In future work, we will explore both higher-order response quantities and determination of complete spectra using the analytic DMRG response approach.

\section{Acknowledgements}

Support is acknowledged from the Cornell Center for Materials Research, the National Science Foundation through
CHE-0645380 and the Department of Energy, Office of Science through Award No. DE-FG02-07ER46432.

\bibliography{new_derivative}

\end{document}